\documentclass[12pt,a4]{article}
\usepackage{amsmath,amssymb,epsfig}
\usepackage{epstopdf}

\usepackage{color}
\input{colordvi.tex}
\def\unit{{\relax{\rm 1\kern-.26em I}}}
\def \InN{{\unit^n_{N}}}
\def\InNc{{\unit^n_{N_c}}}
\def\unitnp{{({\unit'}^{N_c-n}_{N_c})}}

\newcommand{\lsim}{\lesssim}
\newcommand{\gsim}{\gtrsim}

\makeatletter

\renewcommand\section{\@startsection {section}{1}{\z@}%
                                   {-3.5ex \@plus -1ex \@minus -.2ex}%
                                   {2.3ex \@plus.2ex}%
                                   {\normalfont\large\bfseries}}

\renewcommand\subsection{\@startsection{subsection}{2}{\z@}%
                                     {-3.25ex\@plus -1ex \@minus -.2ex}%
                                     {1.5ex \@plus .2ex}%
                                     {\normalfont\normalsize\bfseries}}

\newcount\hour \newcount\minute
\hour=\time \divide \hour by 60
\minute=\time
\def\now{%
\ifnum \hour<13
  \ifnum \hour=0 \advance \hour by 12 \number\hour:\else \number\hour:\fi%
     \ifnum \minute<10 0\fi%
     \number\minute%
\ A.M.%
\else \advance \hour by -12 \number\hour:%
  \ifnum \minute<10 0\fi%
  \number\minute%
  \ P.M.%
\fi%
}

\makeatother

\begin{document}

\baselineskip=18pt  
\numberwithin{equation}{section}  
\allowdisplaybreaks  



%
%


\thispagestyle{empty}

\vspace*{-2cm}
\begin{flushright}
\end{flushright}

\begin{flushright}
ICRR-588-2011-5\\
IPMU11-0093\\
MCTP-11-21\\
SCIPP-11/04
\end{flushright}

\begin{center}

\vspace{1.4cm}

\vspace{0.5cm}
{\bf\Large 
Constraints on Direct Gauge Mediation Models\\with Complex Representations}
\vspace*{1.5cm}

{\bf
Kentaro Hanaki$^{1}$, Masahiro Ibe$^{2,3}$, Yutaka Ookouchi$^{3}$ and Chang-Soon Park$^{4}$} \\
\vspace*{0.5cm}

$^{1}${\it {Michigan Center for Theoretical Physics, \\ University of Michigan, Ann Arbor, MI 48109-1040, USA}}
\vspace{0.1cm}

${}^{2}${\it Institute for Cosmic Ray Research, University of Tokyo, Chiba 277-8582, Japan }
\vspace{0.1cm}

$^{3}${\it Institute for the Physics and Mathematics of the Universe (IPMU),\\ University of Tokyo, Chiba, 277-8583, Japan}\\
\vspace{0.1cm}

$^{4}${\it Santa Cruz Institute for Particle Physics and Department of Physics, \\ University of California, Santa Cruz, CA 95064, USA}\\

\vspace*{0.5cm}

\end{center}

\vspace{1cm} \centerline{\bf Abstract} \vspace*{0.5cm}
We discuss constraints from cosmology on the models with direct gauge mediation
based on the ISS supersymmetry breaking models.
It is a generic feature that the ISS models  possess a $U(1)$ symmetry which is eventually 
spontaneously broken,
when the models are based on the gauge symmetries with complex representations.
We show that the resultant pseudo Nambu-Goldstone boson causes problems
in cosmology, which leads to severe constraints on the models.
We also show that most parameter space of the models can be probed  by observing gravitational waves 
emitted from the cosmic string network when the $U(1)$ symmetry is a gauge symmetry.

\newpage
\setcounter{page}{1} 



\section{Introduction}
The gauge mediated supersymmetry breaking 
models\,\cite{Dine:1981za,Dine:1981gu,Dimopoulos:1982gm,
Affleck:1984xz,Dine:1993yw,Dine:1994vc,Dine:1995ag} 
are one of the most attractive realizations of the phenomenologically
viable minimal supersymmetric standard model (MSSM).
So far, a lot of models of gauge mediation have been constructed,
for the aim to improve the models in view of naturalness or consistency to 
cosmological observations.
We have now many successful models to be examined by experiments.

A downside of models of gauge mediations is, however,
their complexity.
The models are inevitably complicated, 
since we need to connect the supersymmetry breaking sector
and the MSSM sector via gauge interaction of the MSSM.
In this respect, models of direct gauge mediation 
where the MSSM gauge symmetries are embedded into flavor symmetries
of the supersymmetry breaking sector are particularly attractive since
the models require fewer particles and parameters 
(for a precise definition of direct mediation models, see Refs\,\cite{Dine:2007dz}).
Before the discovery of dynamical supersymmetry
breaking in simple supersymmetric gauge theories by Ref.\,\cite{ISS},
the construction of direct gauge mediation models could rely mostly on 
the vector-like dynamical supersymmetry breaking model developed in 
Refs.\,\cite{Izawa:1996pk,Intriligator:1996pu}, the IYIT model.%
\footnote{See Ref.\,\cite{Murayama:1997pb} for the successful direct mediation
model based on this class of the vector-like dynamical symmetry breaking models.}
Since it was shown that vector-like dynamical supersymmetry breaking 
is possible in more generic models by Ref.\,\cite{ISS}, 
many direct gauge mediation models have been constructed based on the ISS supersymmetry
breaking models (See Refs.\,\cite{Dine:2006xt,Ibe:2006rc,KOO,Terning,Strassler} for early applications.).%
\footnote{
Another advantage of this class of models is that the models have
simple realizations in string theory as emphasized in Refs.\,\cite{HoriOoguri,BraneI,BraneII,BraneIII}
(see \cite{ISreview,KOOreview} for reviews).
}

There are two common features shared by the direct gauge mediation models based on the ISS supersymmetry
breaking models. One, which is shared by all the direct gauge mediation models, is the Landau pole problem. 
Since we embed the MSSM gauge groups into flavor groups of the dynamical sector, 
there appear many fields which transform under the MSSM gauge groups. 
These fields contribute to the beta functions of the gauge coupling constants and drive them to Landau poles below the unification scale. 
This problem forces us to take the messenger scale
rather high, while the supersymmetry breaking scale is kept rather low
to realize the MSSM superparticle masses below the TeV range.
The other feature shared is spontaneous symmetry breaking of global symmetries
at the messenger scale. Especially, when the ISS models are based on  gauge symmetries which 
have complex representations such as $SU(N_c)$, the models possess a $U(1)$ global symmetry
which is consistent with the mass terms of the models. We call it $U(1)_B$ symmetry in the rest of the paper.

In this paper, we discuss the implications of these common properties, high messenger scale and spontaneous breaking of $U(1)_B$ symmetry, of the models. First, if the $U(1)_B$ is a global symmetry, we find that the pseudo Nambu-Goldstone boson (PNGB)
resulting from its spontaneous breaking has a lifetime longer than 
the age of the universe.
In this case, the energy density of the coherent oscillation of the PNGB
easily dominates over the energy density of the universe.
As a result, the models are consistent only when 
the reheating temperature of the universe is much less than hundreds GeV.
Then, if the $U(1)_B$ symmetry is gauged, its spontaneous breaking
does not predict the presence of Nambu-Goldstone boson but does generate cosmic strings with 
a tension of the order of the messenger scale.
As we will discuss, most of the parameter space can be probed 
by observing gravitational waves emitted from the cosmic string networks
in future experiments. Since the existence of the $U(1)_B$ symmetry is strongly interrelated to 
the structure of the gauge dynamics of the supersymmetry breaking sector,
the observation of the gravitational wave gives us an important clue
on the structure of the hidden sector.

 The organization of the paper is as follows. 
 In section 2, we discuss the models of direct gauge mediation and spontaneous
breaking of the global symmetries.
 In section 3,  we discuss cosmological implications of the pseudo Nambu-Goldstone bosons
 resulting from spontaneous $U(1)_B$ symmetry breaking.
In section 4, we discuss cosmological implications for the case where $U(1)_B$ symmetry 
is gauged.
The final section is devoted to our conclusions.

\section{Direct gauge mediation}
In this section, we briefly review the models of direct gauge mediation
based on the ISS supersymmetry breaking model.
The Standard Model gauge groups are embedded into the global symmetries
of the supersymmetry breaking sector.
This class of models often involves global symmetries in addition to those identified with the Standard Model gauge groups.
As we will discuss shortly, some of those extra symmetries are spontaneously broken at the scale
of gauge mediation.
The implications of the spontaneous breaking of those symmetries will be discussed
in the following sections.

\subsection{An explicit example}
To be concrete, we discuss generic properties of the models of direct gauge mediation
by considering an explicit model developed in Ref.\,\cite{KOO}, KOO model,  as an example.
The KOO model is based on the original ISS model \cite{ISS}, which consists of an $SU(N_c)$ gauge 
theory with $N_f$ flavors, $Q_i$ and $\bar Q_i$, with a superpotential,
\begin{eqnarray}
\label{eq:mass}
W= \sum_{i,j=1}^{N_f} (m_Q)_{ij} Q_i \bar Q_j\ ,
\end{eqnarray}
where $m_Q$ is a mass matrix. 
In the KOO model,  the mass matrix $m_Q$ is assumed to be diagonal but split into the first $N_f-N_c$ 
components with the eigenvalues being $m_0$ and the remaining $N_c$ components with $\mu_0$, i.e. $m_Q=\mathrm{diag}(m_0,\cdots,m_0,\mu_0,\cdots,\mu_0)$.
We assume that $m_0 \gg \mu_0$.
In this case, we may rewrite the superpotential into, 
\begin{eqnarray}
 W_{\rm mass} = m_0 (Q^I \bar Q_I) + \mu_0 (Q^a \bar Q_a)\ ,
\end{eqnarray}
where $I=1, \cdots, N_f- N_c$ and $a=1,\cdots, N_c$ denote the flavor indices,
while  the color $SU(N_c)$ indices are contracted in $(Q \bar Q)$.

In this model, the maximal global symmetry $SU(N_f)$ is broken into 
$SU(N\equiv N_f-N_c)\times SU(N_c)\times U(1)_{P'}$ by the mass matrix.
Here, $U(1)_{P'}$ denotes a $U(1)$ subgroup of $SU(N_c)$
with the ratio of the charges of $Q^I$ and $Q^a$ being $N_c$ to $N$.
In addition to those global symmetries, the model also possesses 
a $U(1)$ symmetry with a charge assignment $Q(+1)$ and $\bar Q(-1)$,
which we name $U(1)_B$, 
and a $U(1)$ $R$-symmetry with a charge assignment $Q(+1)$ and $\bar Q(+1)$.

To realize the gauge mediation mechanism, we  embed the MSSM
gauge groups into the subgroup of $SU(N)\times SU(N_c)$ global symmetries.
By this, the MSSM sector is 
connected to the dynamical supersymmetry breaking sector in a direct manner. 
To construct a successful model, however, we need to break
the  $R$-symmetry which forbids the gaugino mass terms.
What is even more troublesome in the models based on the ISS models 
is that the gauginos do not obtain masses comparable
to those of the sfermions even if the $U(1)_R$ symmetry is spontaneously broken
as it happens in the model in Ref.\,\cite{IzawaTobe},
and hence, we need to break the R-symmetries explicitly.%
\footnote{
For generic discussions on the gaugino masses, see Refs.\,\cite{KS,GKK}.
See also Refs.\,\cite{SeqI,Izawa:2008ef,AbelKhoze, HanakiOokouchi,NakaiOokouchi} for related discussions.
}

In the KOO model, the comparable gaugino masses to the sfermion masses are achieved
by adding the following term\,\cite{KOO},
\begin{eqnarray}
W_{\rm def}= - \frac{1}{m_X} (Q^I \bar Q_a) (Q^a \bar Q_I)\ ,
\label{eq:higher-d}
\end{eqnarray}
to the superpotential.
Here, $m_X$ denotes a dimensionful parameter.
With this additional term, the $R$ symmetry is explicitly broken.
The remaining global symmetries are 
$SU(N) \times SU(N_c) \times U(1)_{P} \times U(1)_B$.%
\footnote{We define $U(1)_P$ as a linear combination of $U(1)_B$ and $U(1)_{P'}$ that has a trivial action on $Q_I$.}
In the following discussion, 
we assume that $m_X$ is larger than $m$ but much smaller than 
the Planck scale $M_{PL}$.
Such higher dimensional terms can be, for example,  generated 
by integrating out extra massive fields with masses $m_X$ coupling to $(Q_a, Q_I)$
in a renormalizable theory.
We also assume that other dimension four operators such as $(Q^a\bar{Q}_b)(Q_a\bar{Q}^b)$
are suppressed more significantly than those suppressed by $m_X$, i.e., they are suppressed by the Planck scale,%
\footnote{
Such suppressions can be explained if 
we assume an approximate $R$-symmetry with the charge assignment  
$R(Q_I)= R(\bar{Q}_I) = 1$ and $R(Q_a)= R(\bar{Q}_a) = 0$, 
which is explicitly broken only by the $\mu Q_a \bar{Q}^a$ term\,\cite{KOO}.
}
otherwise, the metastability of the vacuum could be spoiled.

To explore the vacuum structure, 
it is useful to consider the magnetic description,
where the model is well described by the composite fields
\begin{equation}
Y^I_{~J} = Q^I \bar Q_J \;,\qquad Z^I_{~a}= Q^I \bar Q_a\;,\qquad \tilde Z^a_{~I} = Q^a \bar Q_I \;,\qquad \Phi^a_{~b} = Q^a \bar Q_b\;.
\end{equation}
In this description, the superpotential is given by
\begin{equation}\label{KOOSUP}
W= h \mathrm{Tr} \left[ m^2 Y + \mu^2 \Phi - \chi Y \tilde{\chi} - \chi Z\tilde{\rho} - \rho \tilde{Z}\tilde{\chi} - \rho \Phi \tilde{\rho} - m_z Z \tilde Z\right]\ ,
\end{equation}
where $\rho$ and $\chi$ are components of magnetic quarks charged under the 
dual $SU(N_f-N_c)$ gauge group.
Here, we have suppressed all the indices of the gauge and global symmetries and defined
\begin{eqnarray}
m^2 \equiv m_0  \Lambda\ , \quad \mu^2 \equiv \mu_0 \Lambda\ ,\quad
m_z \equiv \Lambda^2/m_X \ .
\end{eqnarray}
Throughout this paper, we assume a dimensionful parameter%
\footnote{The dimensionful parameter $\widehat{\Lambda}$ is related to the scale in electric description $\Lambda$ and the one in magnetic description $\tilde{\Lambda}$ by $${\Lambda^{3N_c-N_f}  \tilde{\Lambda}^{3(N_f-N_c)-N_f}=(-1)^{N_f-N_c}\widehat{\Lambda}^{N_f}\ .}$$ }
in magnetic theory to be the same order as $\Lambda$, i.e. $h\equiv {\Lambda / \widehat{\Lambda}}=O(1)$. 
By appropriate phase rotations, we may take parameters $m$, $\mu$ and $m_z$ to be 
real numbers without loss of generality.

As long as the deformation superpotential \eqref{eq:higher-d} is small, the meta-stable vacuum identified in the ISS model still exists.
In addition to the original supersymmetry breaking vacuum, 
this deformation also leads to other meta-stable 
and supersymmetry breaking vacua which are away from the ISS vacuum.
These vacua are given by
\begin{equation}\label{E:nVacConfig}
\begin{split}
&Y^I_{~J} = \frac{\mu^2}{m_z} \left(\InN\right)^I_J\;,\quad\, \, \,\Phi^a_{~b} = \frac{m^2}{m_z} \left(\InNc\right)^a_{~b} + O(m_Z)\ ,\\
&\chi^I_{~J} = m \delta^I_J\;,\,\qquad\qquad \tilde\chi^I_{~J} = m \delta^I_J\ ,\\
&\rho^I_{~a} = \mu \Gamma^I_{~a}\;,\,\qquad \qquad \tilde\rho^a_{~I} = \mu \Gamma^a_{~I}\ ,\\
&Z^I_{~a} = -\frac{m\mu}{m_z}\Gamma^I_{~a}\;,\qquad \tilde Z^a_{~I} = - \frac{m\mu}{m_z}\Gamma^a_{~I}\ ,
\end{split}
\end{equation}
where $n$ is any integer between 0 and $N$.
The matrices $\Gamma^a_{~I}$ and $\Gamma^I_{~a}$ have $1$ in the first $n$ diagonal elements 
and $0$ elsewhere;
\begin{equation}
\Gamma^a_{~I} = \begin{pmatrix} \unit_n & 0_{n\times ({N-n})} \\ 0_{(N_c-n)\times n} & 0_{(N_c-n)\times (N-n)} \end{pmatrix}\;,\qquad
\Gamma^I_{~a} = \begin{pmatrix} \unit_n & 0_{n\times (N_c-n)} \\ 0_{(N-n)\times n} & 0_{(N-n)\times (N_c-n)} \end{pmatrix}\;.
\end{equation}
The matrix $\unit_m$ is an $m\times m$ identity matrix and $\unit^p_q$ is a $q\times q$ matrix whose first $p$ diagonal elements are $1$ and $0$ otherwise. 
The vacuum with $n=0$ corresponds to the vacuum of the ISS model, while the vacua with $n>0$
have lower energy than the ISS vacuum, and the supersymmetry breaking scale of these vacua is $O(\mu^2)$.
As we show in the appendix A, all pseudo-moduli are stabilized by the Coleman-Weinberg potential.

At the vacuum in Eq.\,\eqref{E:nVacConfig}, the 
global symmetries are spontaneously broken into $SU(n) \times SU(N - n) \times SU(N_c -n) \times U(1)^2$ for $n>0$. 
Here, two preserved $U(1)$ symmetries are linear combinations of the following three $U(1)$ symmetries;
\begin{eqnarray}\label{E:U1P}
U(1)_P : \rho \to e^{i \theta_P} \rho\ .
\end{eqnarray}
\begin{eqnarray}\label{E:U1c}
U(1)_c : \rho \to  \left(
\begin{array}{cc}
e^{i \theta_c} {\bf 1}_{n \times n} & 0\\
0 & e^{i \frac{n}{N_c - n}\theta_c} {\bf 1}_{(N_c -n) \times (N_c - n)}\\
\end{array}\right) \rho\ ,
\end{eqnarray}
\begin{eqnarray}\label{E:U1d}
U(1)_d : \rho \to \rho \left(
\begin{array}{cc}
e^{i \theta_d} {\bf 1}_{n \times n} & 0\\
0 & e^{i \frac{n}{N_f -N_c - n}\theta_d} {\bf 1}_{(N_f - N_c -n) \times (N_f -N_c - n)}\;.
\end{array}\right)\ .
\end{eqnarray}
Then one of the preserved symmetries is a combination of $U(1)_P$ and $U(1)_c$ with $\theta_P = - \theta_c$ and the other is a combination of $U(1)_c$ and $U(1)_d$ with $\theta_c = - \theta_d$.
In contrast, at the highest energy vacuum, i.e. the ISS vacuum, 
only a $U(1)$ global symmetry is spontaneously broken
and the resulting preserved symmetry is $SU(N_f - N_c) \times SU(N_c) \times U(1)_P$.
In both cases, one of the $U(1)$ symmetries is spontaneously broken,
which corresponds to the $U(1)_B$ symmetry in the electric description.

Finally, let us embed the MSSM gauge groups
into the flavor symmetry and construct direct gauge mediation.
In this case, the admixtures of the dual quarks and the mesons
play roles of the messengers whose masses are around $m$.
The detailed analysis of the mass matrix of 
the supersymmetry breaking sector including the messengers
is given in the appendix A.
When we embed the MSSM gauge groups into 
$SU(N-n)$, the messenger mass matrix squared of the fermions is given by $A_2$ in \eqref{E:A2}. 
This is nothing but the one appearing in O'Raifeartaigh model of KOO vacua\,\cite{KOO}. 
Thus, just replacing $N_c$, $N$ in the result of \cite{KOO} with $N_c-n$, ${N-n}$ respectively, 
we obtain the leading order of the gaugino masses,
\begin{eqnarray}
m_{\lambda}\simeq {g^2 {(N_c-n)}\over (4\pi)^2}F_\Phi {\partial \over \partial \Phi} {\log\det }M={g^2 {(N_c-n)}\over (4\pi)^2} {h \mu^2 m_z \over m_z \langle \Phi \rangle -m^2}\ .
\end{eqnarray}
Here, $\langle \Phi \rangle $ is determined by the Coleman-Weinberg potential for full O'Raifeartaigh model including the contributions from $A_1$ and $B_1$ (see the appendix A), 
which is proportional to the explicit R-breaking parameter $m_z$, i.e. 
$\langle \Phi \rangle \sim m_z$. 
As a result,  at the leading order of $m_z/ m$, the gaugino masses are given by,
\begin{eqnarray}
m_{\lambda}\simeq -{g^2 {(N_c-n)}\over (4\pi)^2} {h \mu^2 m_z \over  m^2}\left( 1 +{ O}
\left({m_z^2  \over m^2}\right) \right)\ .
\end{eqnarray}
When we embed the MSSM gauge groups into the $SU(N_c-n)$ group, 
gaugino masses are given by
\begin{eqnarray}
m_{\lambda}\simeq -{g^2 {(N-n)}\over (4\pi)^2} {h \mu^2 m_z \over  m^2}\left( 1  +{ O}
\left({m_z^2  \over m^2}\right)\right)\ .
\end{eqnarray}

Notice that if we embed the MSSM gauge groups into the $SU(n)$ symmetry, 
the messenger sector does not include a tachyonic direction anywhere in the moduli space, 
and hence, one expects that gaugino masses vanish\,\cite{KS}. 
In fact, we can show ${\rm det}M=\mu^2 m_z$ by a direct computation,
which indicates that the determinant of the matrix is independent of $\Phi$, 
and hence, gaugino masses vanish.

The sfermion masses are, on the other hand, roughly given by,%
\footnote{
The sfermion squared mass contributions from $\chi$'s 
correspond to the ones discussed in Ref.\,\cite{Green:2010ww,Auzzi:2010mb,Sudano:2010vt},
which tend to be suppressed compared with the other 
contributions from $\rho$'s and $Z$'s.
The detailed analysis on the sfermion squared masses will be given elsewhere.
}
\begin{eqnarray}
 m_{\rm sfermion} \sim \frac{g^2} {4\pi}\frac{\mu^2}{m}\ .
\end{eqnarray}
In order to have comparable gaugino masses to the sfermion masses
in the TeV range,  
the mass parameters should satisfy
\begin{eqnarray}
\label{eq:scales}
\frac{\mu^2}{m} &\sim & 100\,{\rm TeV}\ , \\
m_z &\sim & m\ .
\end{eqnarray}
Here, the second condition is realized when
\begin{eqnarray}
\label{eq:LmX}
\Lambda \sim \sqrt{m m_X}\ .
\end{eqnarray}

As discussed in Ref.\,\cite{KOO}, the Landau pole problem is solved by assuming 
that the messenger scale is higher than $10^{12}$\,GeV.
Besides, the supersymmetry breaking scale is  restricted to be smaller than $10^{9.5}$\,GeV
to avoid unacceptably large flavor violating soft parameters from the gravity mediation effects.
By combining these constraints with Eq.\,\eqref{eq:scales}, we find that 
the model is viable for
\begin{eqnarray}
10^{12}\,{\rm GeV}\lesssim &m&\lesssim 10^{14}\,{\rm GeV}\ , \cr
10^{8.5}\,{\rm GeV}\lesssim &\mu&\lesssim 10^{9.5}\,{\rm GeV}\ .
\end{eqnarray}

Let us summarize the important features of the models with direct
gauge mediation based on the ISS supersymmetry breaking models.
In the KOO model,
the global $U(1)_B$ symmetry is spontaneously broken 
by the expectation values of $\chi$ and $\tilde \chi$ in Eq.\,\eqref{E:nVacConfig}.
Since the messengers obtain masses from the vacuum expectation value of 
$\chi$ and $\tilde\chi$,
the $U(1)_B$ breaking scale coincides with the messenger scale.
Notice that this non-trivial interrelation between the messenger
scale and the $U(1)_B$ breaking scale is rather generic feature of
the models of direct gauge mediation based on the ISS variant models.

In the following discussion, we concentrate on the vacuum with $n=0$,
since it is the vacuum closest to the symmetry enhancement point which can be 
dynamically chosen in the cosmological evolution.%
\footnote{We may straightforwardly extend our analysis to those with $n>0$.}
We also assume the models where the MSSM gauge groups are embedded 
into $SU(N)$ with an assumption $N_c > 10$ to avoid the Landau pole problem
(see \cite{KOO} for details).

\section{Cosmological implication of the PNGB}
As emphasized in the previous section,
spontaneous symmetry breaking of global symmetries
is expected at the messenger scale 
in the direct gauge mediation models.
If the symmetries are exact, the corresponding Nambu-Goldstone bosons are exactly massless.
It is generally believed, however,  that any global symmetries should be broken in gravitational theory.%
\footnote{
This can also be understood by the fact that there is no global symmetry in string theory. See \cite{Banks:2010zn} for a recent discussion.}
That is, the global symmetries are expected to be explicitly broken at least by 
higher dimensional
operators which are suppressed by the Planck scale. 
Therefore the resultant Nambu-Goldstone bosons are pseudo Nambu-Goldstone bosons (PNGBs)
and obtain finite masses.

When the PNGBs are massive but very light, their lifetime can be very long 
and cause problems in the cosmological evolution.
To avoid such problems, the symmetry breaking operators 
should not be suppressed so much so that the resultant PNGBs are heavy enough.
As we will see,  however, the $U(1)_B$ symmetry is highly protected by the gauge 
symmetry and the corresponding PNGB  cannot be heavy enough.%
\footnote{At the vacuum with $n>0$, 
we also have spontaneous breaking of the non-abelian global symmetries.
Those symmetries, however, can be broken even by the tree-level superpotential 
in Eq.\,\eqref{eq:mass}.
Therefore, the PNGBs from the non-abelian global symmetries may 
have masses at the messenger scale.
}
As a result, the PNGB from $U(1)_B$ breaking dominates
over the energy density of universe, which ruins the success of the standard cosmology.

\subsection{PNGB mass}
The distinctive feature of the $U(1)_B$ symmetry is that 
it is highly protected by the $SU(N_c)$ gauge symmetry
of the supersymmetry breaking dynamics.
Even the lowest dimensional operators which explicitly break the $U(1)_B$
symmetry are quite suppressed;
\begin{eqnarray}
 W_B^{\rm (leading)} \sim \frac{C_B}{M_{PL}^{N_c-3}} Q^{N_c}+\frac{ C_{\bar B}}{M_{PL}^{N_c-3}} \tilde{Q}^{N_c}\ ,
\end{eqnarray}
where $C_{B,\bar B}$ are dimensionless coefficients of $O(1)$.
These operators correspond to the baryon and anti-baryon operators, respectively.
By remembering 
that the size of the gauge group $SU(N_c)$ 
is required to be large, $N_c > 10$ (see Ref.\,\cite{KOO} for details), 
we see that these operators are quite suppressed.

In the magnetic dual description, the above baryon and anti-baryon operators are translated into,
\begin{eqnarray}
W_{B}^{\rm (leading)}\sim 
c_B{\Lambda^{2N_c-N_f} \over  M_{PL}^{N_c-3}}\chi^{N_f-N_c} 
+ c_{\bar B}{\Lambda^{2N_c-N_f} \over  M_{PL}^{N_c-3}}\tilde{\chi}^{N_f-N_c} \ ,
\end{eqnarray}
where $c_{B,\bar{B}}$ are again the dimensionless coefficients of $O(1)$.
The dynamical scale $\Lambda$ of the $SU(N_c)$ gauge theory has been incorporated based on the dimensional analysis. 

By combining the superpotential of the KOO model in Eq.\,(\ref{KOOSUP})
with the above $U(1)_B$ breaking operators, we obtain,
\begin{eqnarray}
\label{eq:potentials}
W&=& \bar{m}\Lambda M+ q\tilde{q}M+{\Lambda^2 \over m_{X}}Z\tilde Z
+{\Lambda^2 \over M_{PL}}M^2
+ W_B + \cdots\ ,\nonumber\\
W_{B} &=&W_B^{\rm (leading)} + {\Lambda^{2N_c-N_f+1}\chi^{N_f-N_c} M\over M_{PL}^{N_c-1}}
+ {\Lambda^{2N_c-N_f+1}{\tilde \chi}^{N_f-N_c} M\over M_{PL}^{N_c-1}}+\cdots\ ,
\end{eqnarray}
where the meson field $M$ collectively denotes the meson fields $Y$, $\Phi$ and $Z$,
and the dual quarks, $q$ and $\tilde q$  denote $q={(\chi, \rho)}$ 
and $\tilde q={(\tilde \chi,\tilde \rho)}$.
We have omitted detailed flavor structures of each term.
In the above superpotential, we  included the higher dimensional 
$U(1)_B$ breaking operators which are relevant for the PNGB mass.

The leading contribution to the PNGB mass is from the cross term of the fourth term of $W$ 
and the second and third terms of $W_B$ in Eq.\,(\ref{eq:potentials})
via
${\cal L}\sim |\partial_M W|^2 $, which roughly leads to,
\begin{eqnarray}
\label{eq:mass1}
m_{\rm PNGB}^2 \sim {\Lambda^{2N_c-N_f+3} \over M_{PL}^{N_c}} m^{N_f-N_c-1}\ .
\end{eqnarray}
Here, we have used $\langle{\Phi}\rangle\sim m_z \sim m$.
Notice that the contributions from the second and the third terms of $W$ are vanishing at the leading order since $\langle{Z\chi}\rangle=\langle{\tilde\chi\tilde Z}\rangle = 0$.

\begin{center}
\begin{figure}[tbp]
\begin{minipage}{.5\linewidth}
  \includegraphics[width=.8\linewidth]{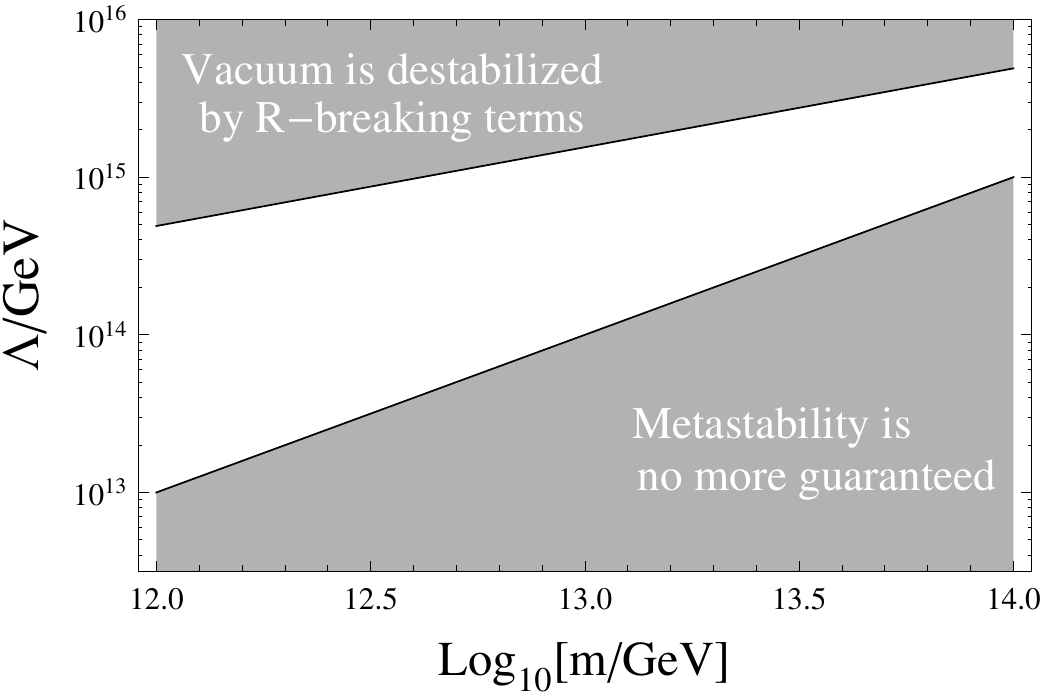}
\end{minipage}
\begin{minipage}{.5\linewidth}
  \includegraphics[width=.8\linewidth]{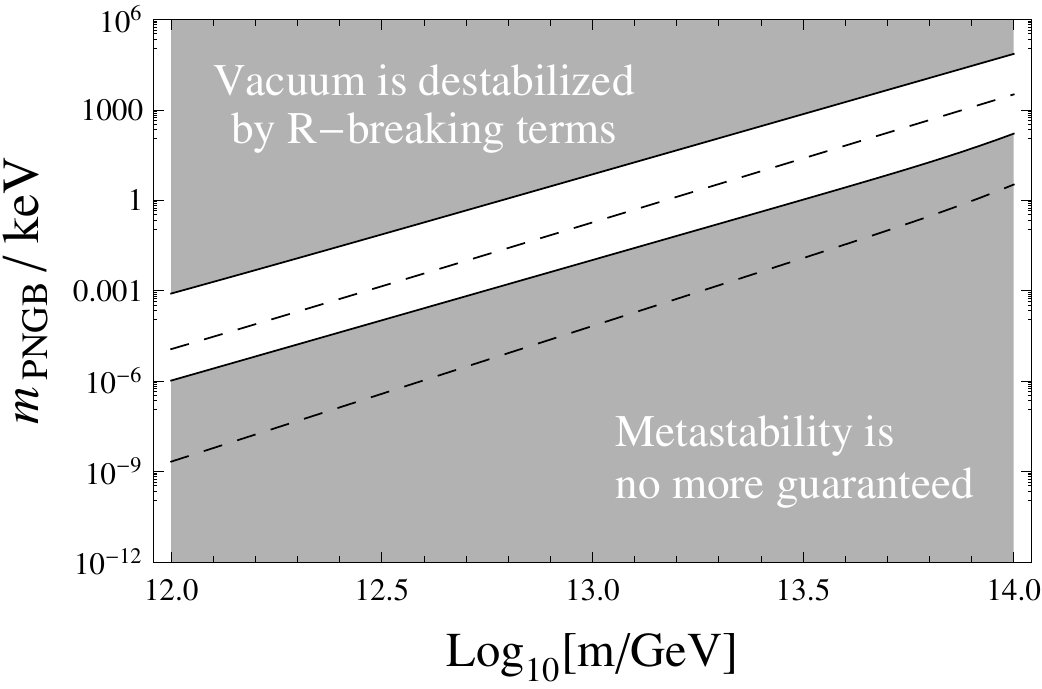}
\end{minipage}
 \caption{\sl \small
Left) The parameter region of the dynamical scale $\Lambda$ (see Eq.\,(\ref{eq:Lconditions})). 
Right) The PNGB mass as a function of $m$.
The upper solid line corresponds to $m_{\rm PNGB}$ for $\Lambda = \Lambda_{\rm max}$ 
for $N_c=10$, where $m_{\rm PNGB}$ is dominated by the one in Eq.\,(\ref{eq:mass1}).
The lower solid line corresponds to $m_{\rm PNGB}$ for for $\Lambda =\Lambda_{\rm min}$
for $N_c=10$, where $m_{\rm PNGB}$ is dominated by the one in Eq.\,(\ref{eq:mass2}). 
The dashed lines denote those values for $N_c =11$.
}
 \label{fig:PNGBmass}
\end{figure}
 \end{center}

Another relevant contribution comes from the $A$-terms of $W_B^{(\rm leading)}$
via the higher dimensional   K\"ahler potential term,
\begin{eqnarray}
\delta K \sim {MM^{\dagger}qq^{\dagger}\over \Lambda^2}\ .
\end{eqnarray}
By plugging $M=m+\theta^2 \mu^2$ and renormalizing the kinetic term of $q$, we obtain the 
A-term of $W_{B}^{(\rm leading)}$,
\begin{eqnarray}
{\cal L} \sim (N_f-N_c)\frac{\mu^2}{\Lambda^2}m\times W_{B}^{(\rm leading)} + h.c.\ ,
\end{eqnarray}
which leads to
\begin{eqnarray}
\label{eq:mass2}
m_{\rm PNGB}^2\sim (N_f - N_c)\mu^2 {\Lambda^{2N_c-N_f-2}\over M_{PL}^{N_c-3}}m^{N_f-N_c-1}\ .
\end{eqnarray}
By comparing Eqs.\,(\ref{eq:mass1}) and \,(\ref{eq:mass2}) with the conditions
in Eq.\,(\ref{eq:scales}), we find that the PNGB mass from the $A$-terms is dominant 
when the dynamical scale $\L$ is relatively low (see Figure\,\ref{fig:PNGBmass} for numerical results).

Before giving the numerical results of the PNGB mass,
let us discuss the constraints on the dynamical scale $\Lambda$.
For generating the gaugino masses comparable to the sfermion masses,
the $R$-breaking mass term,  $m_Z Z\tilde Z$, played an important role.
As we mentioned earlier, the other mass terms such as $\Phi^2$ should be, on the other hand,
much smaller than $m_Z Z\tilde Z$ 
otherwise the metastable supersymmetry breaking vacua 
would be destabilized. 
Thus, we need to require $m_X \ll M_{PL}$.
By remembering that we also require $\Lambda \sim \sqrt{m m_X}$ 
for a successful gauge mediation (see Eq.\,(\ref{eq:LmX})),
this condition leads to the upper bound on the dynamical scale for a given messenger scale.

The dynamical scale $\Lambda$ is also constrained from below,  $\Lambda \gg m$, otherwise
the metastability of the supersymmetry breaking vacua could not be analyzed.
That is, the incalculable higher dimensional terms of the K\"ahler potential 
of the pseudo-flat direction such as,
\begin{eqnarray}
 K \sim \frac{|\Phi|^4}{\Lambda^2}\ ,
\end{eqnarray}
become comparable to the radiatively generated K\"ahler potentials via the dual quark
interactions.

In Fig.\,\ref{fig:PNGBmass}, we show the upper and lower limits on $\Lambda$ as a function 
of $m$ which are defined by,
\begin{eqnarray}\label{eq:Lconditions}
   \Lambda_{\rm max}  = \sqrt{0.1\times M_{PL} m}\ ,\quad \Lambda_{\rm min} = 10\times m \ .
\end{eqnarray}
The figure shows that the dynamical scale is in a range of $10^{13-16}$\,GeV.
In the right panel of Fig.\,\ref{fig:PNGBmass}, we show the PNGB mass for 
a given messenger scale.
The figure shows that the PNGB mass is quite suppressed, and for example,
\begin{eqnarray}
 m_{\rm PNGB} \ll O(1)\, {\rm keV}\ ,
\end{eqnarray}
for $m \sim 10^{13}$\,GeV.
As we will see shortly, the energy density of the PNGB dominates over 
the energy density of the universe due to the long lifetime, although
it is very light.

Before closing this section, 
let us discuss the masses of the fermionic and the scalar superpartners of the PNGB.
Since the $U(1)_B$ symmetry is broken at the  energy scale higher than the supersymmetry breaking scale, these partners are also expected to be light. 
Unlike the PNGB, however, they obtain masses by the supersymmetry breaking effects. 
To discuss their masses, it is useful to define the PNGB chiral supermultiplet ${\cal A}$ by,
\begin{eqnarray}
 \chi \sim m \times e^{\frac{\cal A}{m}}\ , \quad \tilde\chi \sim m \times e^{\frac{-\cal A}{m}}\ . 
\end{eqnarray}
The partners of the PNGB obtain the masses via the K\"ahler potential,
\begin{eqnarray}
\label{eq:PNGBK}
 K \sim \frac{h^2}{16\pi^2} \frac{\Phi^\dagger\Phi}{m^2} ({\cal A} +{\cal A}^\dagger )^2 + \cdots \ ,
\end{eqnarray}
which is radiatively generated by the $\chi-Z-\rho$ interactions.
By plugging $\Phi \simeq m + \mu^2\theta^2$ into the above K\"ahler potential, 
we obtain the masses of the superpartners  of the PNGB;
\begin{eqnarray}
m_{\rm fermion} \sim \frac{h^2}{4\pi^2} \frac{\mu^2}{m}\ ,  \quad
m_{\rm scalar} \sim \frac{h}{4\pi} \frac{\mu^2}{m}\ .
\end{eqnarray}
Therefore, they are typically heavier than the superparticles in the MSSM unlike the PNGB of the $U(1)_B$ symmetry.


\subsection{PNGB decay width}
As pointed out in Ref.\,\cite{Banks}, discrete symmetries such as charge conjugation symmetry and CP-symmetry play important roles to determine 
the interaction between the PNGB and the MSSM sector.

Let us first consider a charge conjugation transformation defined by the exchange,
\begin{eqnarray}
\label{eq:conjugate}
 Q \leftrightarrow \bar{Q}\ ,
\end{eqnarray}
with appropriate charge conjugations of the gauge fields.
The PNGB is odd under this transformation.
This charge conjugation is a good symmetry when we neglect  the matter 
fields in the MSSM sector.
Thus, as long as we integrate out only the messenger fields,
the effective interactions of the PNGB respect this symmetry.
As a result, the lowest dimensional operators which are relevant for 
the decay of the PNGB into the MSSM gauge fields 
require at least three gauge fields which are roughly given by\,\cite{DolgovI,DolgovII},
\begin{eqnarray}
{\cal L}_{\rm eff}\sim \frac{g_{SM}^3}{4\pi}{1 \over m^7} (D_{\rho} F_{\alpha \beta} )(D^{\beta} F_{\sigma \tau})( D^{\rho} D^{\alpha} F^{\sigma \tau}) {\cal P} \ ,
\end{eqnarray}
where ${\cal P}$ denotes the PNGB, $F_{\mu\nu}$ the gauge field strengths of the Standard Model,
and $D^\mu$ the appropriate covariant derivatives. 
We collectively denoted the corresponding gauge coupling constants by $g_{SM}$.
Therefore, the decay width of the PNGB into photons through this operator is
negligibly small.

Once we integrate the MSSM sector as well, the above non-renormalizable operator
could lead to the effective interactions between the PNGB and the Standard Model fermions, $f$, 
at the three-loop level,
which are at most given by,
\begin{eqnarray}
 {\cal L}_{\rm eff} \sim \left(\frac{\alpha_{SM}}{4\pi}\right)^3\frac{1}{m} \partial_\mu {\cal P}
 (f^{\dagger} \sigma^\mu f + \cdots )\ .
\end{eqnarray}
Here,  the gauge coupling constants 
are estimated at the messenger scale.
The decay widths of the PNGB into the fermions (i.e. the electrons or neutrinos)
through these operators are quite suppressed, and hence, the dominant process should 
be the one into photons through the diagrams with one more loop.
As a result, the decay width is at most of the order of
\begin{eqnarray}
 \Gamma \sim  \left(\frac{\alpha_{SM}}{4\pi}\right)^8 \frac{m_{\rm PNGB}^3}{m^2}
 \sim 10^{-60}\,{\rm GeV}\times \left(\frac{m_{\rm PNGB}}{1\,\rm keV} \right)^3
 \left(\frac{10^{13}\,\rm GeV}{m} \right)^2
 \ . 
\end{eqnarray}
Therefore, we find that the lifetime of the PNGB is much longer than the age of the universe.%
\footnote{The scalar and the fermionic partners of the PNGB decay immediately into a pair of the PNGBs
or a pair of the PNGB and the gravitino respectively via the operator in Eq.\,(\ref{eq:PNGBK}).
}

The PNGB may also decay via the interaction terms which explicitly break the $U(1)_B$ symmetry
in Eq.\,(\ref{eq:potentials}).
From the explicit breaking term $W_B^{\rm (leading)}$, the messenger fields 
obtain an effective mass term which depends on the PNGB,
\begin{eqnarray}
\delta m_{q} \sim C_B {\Lambda^{2N_c-N_f} \over M_{PL}^{N_c-3}} m^{N_f-N_c-2}e^{i{\cal P}/m} \ .
\end{eqnarray}
Thus, after integrating out messengers, we obtain an interaction terms,
\begin{eqnarray}
{1\over 16\pi^2}\int d \theta^2   \log M_{\rm mess} W_{\alpha}W^{\alpha} \sim  {1\over 16\pi^2} ( c_B-c_{\bar{B} })    {\Lambda^{2N_c-N_f} \over M_{PL}^{N_c-3} } m^{N_f-N_c-4}{\cal P}F_{\mu \nu } \widetilde{F}^{\mu \nu} \ .
\end{eqnarray}
Notice that these operators vanish for $c_B = c_{\bar B}$ since
 the charge conjugation is a good symmetry in this limit.
As a result, the PNGB decays into two photons via these operators with the decay width, 
\begin{eqnarray}
\Gamma \simeq \left( {1\over 16\pi} {m_{\rm PNGB}^3 \over M_{\rm eff}^2} \right)\ ,
\end{eqnarray}
where we have defined 
\begin{eqnarray}
M_{\rm eff}=\frac{16\pi^2}{c_B-c_{\bar B}}M_{PL} \left({M_{PL}\over \Lambda} \right)^{2N_c-N_f} \left( {M_{PL}\over m}\right)^{N_f-N_c-4} \gg M_{PL} \ .
\end{eqnarray}
The last inequality is obtained by remembering $N_f-N_c > 5$ and $2N_c - N_f > 0$.
As a result,  the decay of the PNGB via the $U(1)_B$ breaking operators
is subdominant, and hence, the lifetime of the PNGB is much longer than the age of the universe.

\subsection{PNGB dominated universe}
Now we are ready to study energy density of a coherent oscillation of
the PNGB in the history of cosmology.
During inflation, the $U(1)_B$ symmetry is expected to be restored by
the inflation dynamics if the Hubble scale during inflation is larger than $m$,
since $\chi$'s are expected to obtain effective masses around the symmetry enhancement point, 
i.e. $\chi = \tilde \chi = 0$.
In this case, the initial amplitude of the coherent oscillation of the PNGB from its true minima%
\footnote{Let us remember that the $U(1)_B$ symmetry is explicitly broken.}
after inflation is $O(m)$.
If the Hubble scale during inflation is smaller than $m$, the $U(1)_B$ 
symmetry is not restored during inflation.
Even in this case, the initial amplitude of the coherent oscillation of the PNGB after inflation 
is again expected to be $O(m)$,
since the PNGB has a very flat potential on which
the initial condition is randomly distributed.
Thus, as long as the Hubble scale during inflation is much larger than $m_{\rm PNGB}$,
the initial amplitude of the coherent oscillation of the PNGB  after inflation is expected to be $O(m)$.
Thus,  the initial energy density of the oscillation of the PNGB after inflation  is expected to be,
\begin{eqnarray}
\label{eq:PNGBA}
 \rho_{\rm PNGB} \sim m_{\rm PNGB}^2 m^2\ .
\end{eqnarray}
In the followings, we discuss the evolution of the energy density of the PNGB after inflation
with this initial condition.

\begin{figure}[t]
\begin{center}
  \includegraphics[width=.5\linewidth]{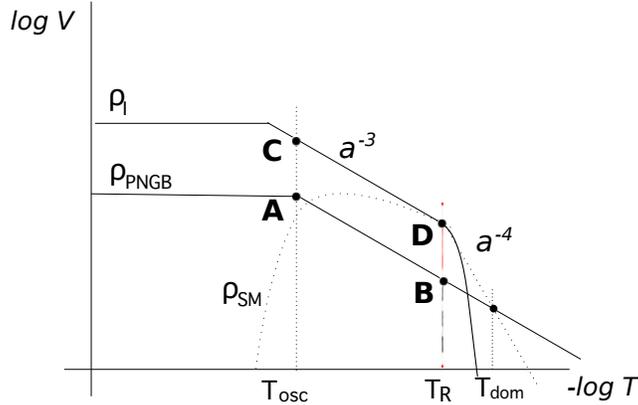}
 \vspace{-.5cm}
\caption{\sl \small The evolution of the energy densities as the universe cools down. 
The two solid lines represent the density for inflatons and that for pseudo-Nambu-Goldstone bosons, respectively, and the dotted line represents that for Standard Model matters. To avoid spoiling the successful Standard Cosmology, 
the energy density, $\rho_{PNGB}$, should not exceed $\rho_{SM}$ at the temperature above 
$T_{eq}=O(1)$\,eV (see text).}
\label{fig:evolution}
\end{center}
\end{figure}

First, let us assume that the reheating temperature of the universe after inflation
is lower than 
\begin{eqnarray}
T_{\rm osc}  \sim \sqrt{m_{\rm PNGB}M_{PL}}\ .
\end{eqnarray}
In this case, the PNGB starts oscillating around its minimum when the Hubble
parameter becomes lower than
\begin{eqnarray}
\label{eq:Hosc}
H < H_{\rm osc}  \sim m_{\rm PNGB}\ ,
\end{eqnarray}
which is before the radiation dominated era (see Fig.\,\ref{fig:evolution}).
Before the universe enters into the radiation dominated era, 
both the inflaton and the PNGB oscillate, and hence,
their energy densities scales in the same way by the cosmic expansion.
Thus, at the beginning of the radiation dominated era, the energy density of the PNGB at the reheating time is given by,
\begin{eqnarray}
 \rho_{\rm PNGB}^{(B)}  \simeq  \rho_{\rm PNGB}^{(A)} \times \frac{\rho_{I}^{(D)}}
 {\rho_{I}^{(C)}} \simeq {T_R^4}\frac{m^2}{M_{PL}^2}\ .
\end{eqnarray}
Here, $\rho_{I}$ denotes the energy density of the inflaton, $T_R$ 
the reheating temperature, and
the superscripts of the energy densities correspond 
to the ones in Fig.\,\ref{fig:evolution}.
In the final expression, we have used Eq.\,(\ref{eq:PNGBA}) and 
\begin{eqnarray}
\rho_{I}^{(C)} &\sim& m_{\rm PNGB}^2 M_{PL}^2\ ,\cr
\rho_{I}^{(D)} &\sim& T_R^4\ ,
\end{eqnarray}
where the first equality is obtained from Eq.\,(\ref{eq:Hosc}) and the
second one represents the energy density conservation at the reheating time.

During the radiation dominated era, the energy density of the radiation scales
by the fourth power of the inverse of the scalar factor $a$, while the
PNGB energy density scales by the third power of the inverse of $a$.
Thus, by remembering that the cosmic temperature scales the inverse  of $a$,
we find that the PNGB energy density dominates the energy density
when the temperature becomes lower than,
\begin{eqnarray}
    T_{\rm dom} \sim T_R \times \frac{\rho_{\rm PNGB}^{(B)}}{\rho_R^{(D)}} \sim T_R \times\frac{m^2}{M_{PL}^2}\ .
\end{eqnarray}
Since the lifetime of the PNGB is longer than the age of the universe,
the domination temperature $T_{\rm dom}$ should be much lower than the 
temperature of the matter-radiation equality temperature of the Standard Cosmology, $T_{\rm eq} = O(1)$\,eV.
Thus, for the successful cosmology, the reheating temperature is severely restricted to 
\begin{eqnarray}
T_R \sim T_{\rm dom} \times \frac{M_{PL}^2}{m^2} \ll T_{\rm eq} \times \frac{M_{PL}^2}{m^2}
\sim O(100)\,{\rm GeV}\times \left(\frac{10^{13}\,{\rm GeV}}{m}\right)^2\ .
\end{eqnarray}
This constraint is rather severe when we consider  baryogenesis mechanisms which 
require very high reheating temperature such as the thermal leptogenesis mechanism\,\cite{Fukugita:1986hr}.

We may do a similar analysis for $T_{\rm osc} < T_R$.
In this case, the PNGB oscillation dominates over the energy density when 
the cosmic temperature becomes lower than 
\begin{eqnarray}
 T_{\rm dom}^{\prime} \sim T_{\rm osc} \times\frac{m^2}{M_{PL}^2}\ .
\end{eqnarray}
Thus, by requiring $T_{\rm dom}^\prime \ll T_{\rm eq}$, we find that 
\begin{eqnarray}
 m_{\rm PNGB} \ll 10^{-10}\,{\rm keV} \times\left(\frac{10^{13}\,\rm GeV}{m}\right)^4\ .
\end{eqnarray}
By comparing the PNGB mass in Fig.\,\ref{fig:PNGBmass}, we find 
that this condition is difficult to be satisfied without causing the Landau pole problem.

\section{Gauged $U(1)_B$ symmetry}
A simple way out of the above discussion is the gauged $U(1)_B$ symmetry.
When the $U(1)_B$ symmetry is a gauge symmetry, its spontaneous breaking
does not lead to the Nambu-Goldstone boson, and the cosmological constraints discussed above
is not applicable.
Instead, it leads to another cosmologically interesting object, the stable cosmic string.
In this section, we discuss the implications of the cosmic string 
in the direct gauge mediation models based on the deformed ISS models.

\subsection{Classification of vortex solutions}
Here, we show that there are two types of vortices in the deformed ISS-model
on which the KOO gauge mediation model bases.%
\footnote{
Topological defects in the ISS-metastable vacuum were firstly studied in Ref.\,\cite{EHT}, 
 where the semi-local vortex solution were explicitly constructed. }
The solutions corresponding to the topological defects have finite energy, and hence, 
the configuration of the solutions decay rapidly enough as the spatial radius increases.
As is well known, this condition is satisfied when the scalar fields responsible for  spontaneous symmetry breaking take values on the same gauge orbit at infinity (see Ref.\,\cite{TongII} for a review.)
For example, the one dimensional topological defect, i.e. the vortex,  corresponds to a map 
from the spatial infinity of two-dimensional space to the gauge orbit 
which defines the winding number $\pi_1(G_{\rm local}/H_{\rm local})$
(see also the Appendix B). 
The homotopy group, $\pi_1(G_{\rm local}/H_{\rm local})$,  can be calculated by 
using the exact sequence,
\begin{eqnarray}
&&\pi_2(G)=0  \xrightarrow{f_1} \pi_2(G/H) \xrightarrow{f_2} \pi_1 (H)  \xrightarrow{f_3} \pi_1 (G) \xrightarrow{f_{4} } \pi_1 (G/H)   \xrightarrow{f_{5} } \pi_0(H)=0 \ , \nonumber \\
&&\hspace{5cm}{\rm Ker}f_{k+1}={\rm Im}f_k\ .
\end{eqnarray}

Let us start with the highest energy vacuum where the only non-vanishing dual quarks in \eqref{KOOSUP} are $\chi^I_{~J}=\tilde{\chi}^I_{~J}=m \delta^I_{~J}$ up to symmetry transformations. In this case, symmetry breaking occurs as follows:
$${
G \equiv SU(N)_{\rm color}\times SU(N)\times SU(N_c) \times U(1)_B \times U(1)_P \to H\equiv SU(N)_{\rm diag}\times SU(N_c) \times U(1)_P\ .
}$$
Since $U(1)_P$ acts trivially on this metastable vacuum, non-trivial contribution comes only from the gauged $U(1)_B$ breaking,
$${
\pi_1\left( {G_{\rm local}\over H_{\rm local}} \right)=\pi_1\left( { SU(N)\times U(1)_B \over {\bf 1}} \right)={\bf Z}\ .
}$$
Therefore, we find that there is at least one finite energy vortex. 

The vortex solutions at the lower energy vacua are more interesting. 
Gauging only $U(1)_B$ is very similar to the case of the highest KOO vacuum. So we here consider a vortex when both $U(1)_B$ and $U(1)_P$ are gauged. The fundamental group is 
$${
\pi_1\left( {SU(N)\times U(1)_B \times U(1)_P \over {\bf 1}} \right)={\bf Z}\times {\bf Z}\ . 
}$$
So there should be {\it two} types of finite vortices. 
As we discuss in the Appendix B, one of the vortices corresponds 
to the so-called semi-local vortex.
In the following analysis, we again concentrate on the case of the highest energy vacuum, $n=0$.

\subsection{Cosmic strings}
Now let us discuss the cosmological implications of the vortex solutions.
Since the $U(1)_B$ symmetry is gauged and $\pi_1(G_{\rm local}/H_{\rm local})={\bf Z}$, 
dynamical generation of the cosmic strings with a tension
\begin{eqnarray}
 \mu_T \simeq \pi^2\, m
 \end{eqnarray}
is expected at the phase transition of the $U(1)_B$ symmetry. 
Such phase transition occurs when the Hubble scale 
or the temperature of the universe become lower than the 
symmetry breaking scale $m$.
At the phase transition, typically one cosmic string is generated in a Hubble volume\,\cite{Kibble}.%
\footnote{
This estimation represents that the causality does not permit an exchange of information beyond the horizon scale. 
In Ref.\,\cite{Zurek}, however, it is pointed out that if the system cools quickly a field value can be different within a causally connected region. 
As recently emphasized in Ref.\,\cite{murayamaDM}, this Kibble-Zurek mechanism provides a substantially larger abundance of topological defects from phase transitions in the early universe than the original estimation by Kibble.
Notice however that the energy density of the cosmic strings  on its
simple scaling low (see below) is insensitive to the initial energy density of the cosmic strings.
}

First, let us consider the case with $m > H_{\rm inf},  T_{H}$,
where $H_{\rm inf}$ is the Hubble scale during inflation and $T_H$ the highest 
temperature afte inflation.%
\footnote{
The highest temperature $T_{H}$ during inflation is given by $T_{H} \simeq (T_{R}^2 H_{\rm inf} M_{PL})^{1/4}$.
}
In this case, the cosmic strings are diluted away by the inflation even if 
they are generated at the phase transition before the inflation.
Therefore, we find that the models with direct gauge mediation based on the deformed ISS is consistent with cosmology if the $U(1)_B$  symmetry is a gauge symmetry,
since cosmic strings do not leave any visible effects after inflation.
Since the messenger scale $m$ is above $10^{12}$\,GeV,
the condition $m>T_H$ is still consistent with thermal leptogenesis\,\cite{Fukugita:1986hr}.

Next, let us consider the cases with ${\rm Max}\, [H_{\rm inf},T_{H}] \gsim  m$  
where the cosmic strings are not inflated away. 
In these cases, the phase transition takes place after the end of inflation,
and the produced cosmic strings survive until now.

Once  very long cosmic strings are formed at the phase transition, their energy density 
is naively expected to scale as $\rho_{\rm str}\sim \mu_T H_*^2 a^{-2}$, where $H_*$
is the Hubble parameter at the phase transition.
If this is the case, the energy density of the cosmic strings eventually dominates
over the energy density of the universe unless the tension $\mu_T$ is negligibly small.
However, the long cosmic strings break into loops through the reconnection processes of the strings.
The produced loops subsequently disappear by emitting gravitational waves.
In this way, the energy density of the string finally reaches the simple scaling,
$ \rho_{\rm str} \propto \mu_T H^2$, despite the very complicated system
of the string network (see \cite{Book} for details). 
As a result, the energy density of the stable cosmic string
is always subdominant at any time as long as the tension $\mu_T$
is much smaller than the Planck scale.

One of the striking effects of the cosmic string is its contributions to the anisotropy
of the cosmic microwave background radiation (CMBR).
Recent CMBR observations from WMAP and SDSS experiments
 have placed a limit on the tension of the cosmic string, 
$G\mu_T < 3.5 \times 10^{-7}$\,\cite{Wyman:2005tu}
where $G$ is the Newton constant, $G \simeq(1.2 \times 10^{19}\,{\rm GeV})^2$.
In terms of the messenger scales, the above constraint corresponds to
\begin{eqnarray}
  m\lsim 10^{15.2}\, {\rm GeV} \ .
\end{eqnarray}
Therefore, the messenger scale we are interested in is consistent with the constraints
from the anisotropy of the cosmic background.

Another striking signature from the cosmic strings is the gravitational radiation
produced by the string loops\,\cite{VilenkinGW,Book}.
The most stringent constraint on the cosmic strings through the gravitational wave observation
is the one coming from the pulsar timing tests.
According to Ref.\,\cite{DePies:2007bm}, the tension of the cosmic string 
is constrained to be $G\mu_T < 10^{-9}$ by the pulsar timing limits.%
\footnote{
This value is the result when we assume that the typical length of the newly produced 
loops at the Hubble scale $H$ is about $\ell_{\rm loop} \sim 0.1\times H^{-1}$.
This length at the production is suggested by numerical simulations
done in\,\cite{Vanchurin:2005pa,Ringeval:2005kr,Martins:2005es}.
If the typical length at the production is much shorter,
the pulsar timing test is not sensitive to the gravitational wave from the loops of the cosmic strings.
In such cases, the gravitational waves from the infinite strings
become important at the frequency of the gravitational wave around $10^{-8}$\,Hz 
to which the pulsar timing test is sensitive\,\cite{Kawasaki:2010yi}.
}
Thus, we find that the messenger scale of the direct gauge mediation
is also consistent with the constraint obtained from the pulsar timing limits.

The future experiments of the gravitational wave observation will be much more sensitive to the gravitational waves from the cosmic strings.
As discussed in Ref.\,\cite{DePies:2007bm}, we may probe the gravitational
waves from the cosmic strings with the tension down to  $G\mu_T\simeq 10^{-16}$
by the sensitivity of the LISA experiment for example.%
\footnote{This number is again for $\ell_{\rm loop} \sim 0.1\times H^{-1}$ at the production.
See Ref.\,\cite{DePies:2007bm} for details.}
Thus, the future gravitational wave experiments cover most of the messenger scale,
\begin{eqnarray}
  m > 10^{10.5}\,{\rm GeV}\ .
\end{eqnarray}
Therefore, if the phase transition of the $U(1)_B$ takes place after the end of inflation,
the models based on the deformed ISS can be tested by observing the gravitational wave.%

Finally, let us comment that the Hubble scale during inflation can be determined experimentally 
by the tensor to scalar ratio $r$ of the cosmic microwave background,
since they are related by $H_{\rm inf} \simeq 10^{14}\times r^{1/2}$\,GeV.
The COrE experiment\,\cite{Collaboration:2011ck}, for example, 
aims to measure the ratio down to $10^{-3}$ which
corresponds to $H_{\rm inf} \simeq 10^{12-13}$\,GeV.
In this way, we can indirectly test one of the conditions of the occurrence of the phase transition after
the end of inflation, i.e. $H_{\rm inf} > m$, via the observation of the tensor mode
of the cosmic microwave background.

\section{Conclusions}
In this paper, we discussed cosmological implications 
of the models with direct gauge mediation based on the deformed ISS models.
The generic properties of this class of models is that some of the global symmetries 
are spontaneously broken at the messenger scale.
Especially, we investigated the implications of  spontaneous 
breaking of the $U(1)_B$ symmetry which is an intrinsic feature 
of the ISS models based on the gauge symmetries with complex representations. 

When the $U(1)_B$ symmetry is a global symmetry, it is expected to be 
broken explicitly by gravitational interactions. 
In the ISS models, however, the $U(1)_B$ symmetry is highly protected 
by gauge symmetry of the dynamics of supersymmetry breaking. 
Due to such protection, the mass of the resultant PNGB is
very small and its lifetime is longer than the age of the universe.
The energy density of the coherent oscillation of the PNGB
easily dominates over the energy density of the universe.
As a result, we found that the models with the global $U(1)_B$ symmetry are consistent only when 
the reheating temperature of the universe is much less than hundreds GeV.
This constraint is rather severe when we discuss baryogenesis mechanisms
such as leptogenesis which requires the reheating temperature higher than 
$T_R \gsim 10^9$\,GeV\,\cite{Fukugita:1986hr}.

When the $U(1)_B$ symmetry is a gauge symmetry, on the other hand,
its spontaneous breaking predicts the cosmic strings instead of the Nambu-Goldstone boson.
We found that the messenger scale of our interest, i.e. $m=O(10^{12-14})\,$GeV,
is consistent with the cosmic microwave background constraints and the pulser 
timing limits even if the phase transition occurs after inflation.
Furthermore, we also found that most of the parameter space can be tested 
by observing the gravitational waves from the cosmic string networks
at the future experiments. 

Finally, let us emphasize again that the existence of the $U(1)_B$ symmetry is strongly interrelated to 
the structure of gauge dynamics of the supersymmetry breaking sector.
Thus, the detectability of the remnant of spontaneous breaking of the $U(1)_B$
symmetry allows us to get a glimpse of the structure of the 
hidden supersymmetry breaking sector through the observation of the gravitational waves.
The $U(1)_B$ symmetry is, however, not expected in the IYIT supersymmetry breaking models 
since they are based on the gauge theories with real representations. 
The supersymmetry breaking vacuum of the IYIT model is stable, while that of the ISS model is metastable.
Therefore, by investigating the implications of the $U(1)_B$ symmetry breaking, 
we will not only be able to put constraints on the model parameters, but also might get a hint 
on a speculative question; whether we are on the metastable vacuum or not.

\section*{Acknowledgments}
We would like to thank B. Batell, M. Dine, M. Eto, K. Hashimoto, R. Kitano, Y. Nakai, D. Shih, S. Terashima and T. Watari for discussions and especially K. Ohashi, D. Green, T. Vachaspati and T. Yanagida for useful discussions and suggestions. 
MI thank K. Miyamoto for discussions on cosmic strings.
MI and CP are grateful to the Perimeter Institute for Theoretical Physics for their hospitality. 
YO would like to thank the Institute for Advanced Study and Michigan Center for Theoretical Physics for their hospitality. KH's work was supported in part by the US Department of Energy under grant DE-FG02-95ER40899. YO's research is supported by World Premier International Research Center Initiative (WPI Initiative), MEXT, Japan. CP is also supported in part by DOE grant DE-FG02-04ER41286.


\appendix
\setcounter{equation}{0}
\renewcommand{\theequation}{A.\arabic{equation}}

\section*{Appendix A \, Stability of lower vacua}

In this appendix, we study the stability of the lower energy uplifted vacua in the model studied in \cite{KOO} by calculating the Coleman-Weinberg potential. The supersymmetry condition for the superpotential \eqref{KOOSUP} $\partial W=0$ becomes
\begin{equation}\label{E:KOOSUSYCondition}
\begin{split}
&\mu^2 - \tilde\rho \rho =0\ , \\
&m^2 - \tilde\chi\chi=0\ ,\\
&Y\tilde \chi + Z\tilde\rho=0\;,\qquad \chi Y + \rho\tilde Z =0\ ,\\
&\tilde\rho \chi + m_z \tilde Z =0\;,\qquad \tilde\chi \rho + m_z Z=0\, \\
&\tilde Z\tilde \chi +\Phi \tilde\rho =0\;,\qquad \chi Z + \rho \Phi =0\;.
\end{split}
\end{equation}
Except the first equation, all equations are satisfied, and we obtain supersymmetry breaking vacua.
Note that there are flat directions.
This can be seen by noting that \eqref{E:KOOSUSYCondition} is still satisfied by the change of fields
\begin{equation}
\begin{split}
\chi\rightarrow A_N^{-1} \chi B_N \;,&\qquad \tilde\chi\rightarrow B_N^{-1} \tilde\chi A_N\ ,\\
\rho\rightarrow A_N^{-1} \rho C_{N_c} \;,&\qquad \tilde\rho\rightarrow C_{N_c}^{-1} \tilde\rho A_N\ ,\\
Z\rightarrow B_N^{-1} Z C_{N_c} \;,&\qquad \tilde Z\rightarrow C_{N_c}^{-1} \tilde Z B_N\ , \\
\Phi\rightarrow C_{N_c}^{-1} \Phi C_{N_c} \;,&\qquad Y\rightarrow B_N^{-1} Y B_N\;,
\end{split}
\end{equation}
where $A,B,C$ denote invertible matrices whose sizes are shown in subscripts.
When $A,B,C$ are especially special unitary matrices, they generate Goldstone modes.
This explains half of the possible non-trivial modes.
The other half are pseudo-moduli and the quantum fluctuation may develop a non-trivial Coleman-Weinberg potential.

It suffices to consider the following background fields to obtain the Coleman-Weinberg potential.
\begin{equation}
\begin{split}
&Y^I_{~J} = \frac{\mu^2}{m_z} \left(\InN\right)^I_J\;,\qquad \Phi^a_{~b} = \frac{m^2}{m_z} \left(\InNc\right)^a_{~b}+\gamma \unitnp^a_{~b}\ ,\\
&\chi^I_{~J} = \alpha m \delta^I_J\;,\qquad \tilde\chi^I_{~J} = \alpha^{-1} m \delta^I_J\ , \\
&\rho^I_{~a} = \alpha\beta\mu \Gamma^I_{~a}\;,\qquad \tilde\rho^a_{~I} = \alpha^{-1}\beta^{-1}\mu \Gamma^a_{~I}\ ,\\
&Z^I_{~a} = -\frac{\beta m\mu}{m_z}\Gamma^I_{~a}\;,\qquad \tilde Z^a_{~I} = - \frac{\beta^{-1}m\mu}{m_z}\Gamma^a_{~I}\;,
\end{split}
\end{equation}
where $\alpha$, $\beta$ and $\gamma$ are pseudo moduli.
$\unitnp^a_{~b}$ is an $N_c\times N_c$ matrix whose lower $N_c-n$ diagonal components are 1.
The boson and fermion mass matrices are given by
\begin{equation}
m_B^2 = \begin{pmatrix} W^{\dagger ac} W_{cb} & W^{\dagger abc}W_c \\ W_{abc} W^{\dagger c} & W_{ac} W^{\dagger cb}\end{pmatrix}\;,\qquad
m_F^2 = \begin{pmatrix} W^{\dagger ac} W_{cb} & 0 \\ 0 & W_{ac} W^{\dagger cb}\end{pmatrix}\;.
\end{equation}
From this, the Coleman-Weinberg is computed by using the formula
\begin{equation}\label{E:CWpotential}
V_{eff} = \frac{1}{64\pi^2} \left(\mathrm{Tr}\; m_B^4 \log \frac{m_B^2}{\Lambda^2} - \mathrm{Tr}\; m_F^4 \log \frac{m_F^2}{\Lambda^2}\right)\;.
\end{equation}
Components that are not directly coupled to $W_{\rho\tilde \rho \Phi} W^{\dagger \Phi}$ do not contribute to the Coleman-Weinberg potential due to the boson-fermion cancellation.
Among the off-diagonal components $W_{abc} W^{\dagger c}$ of the boson mass matrix, the only non-vanishing component is
\begin{equation}
W_{\rho^I_a\Phi^b_c \tilde\rho^d_J}W^{\dagger \Phi^a_b} = -\mu^2 \delta^J_I \left[\delta^a_d - \InNc^a_d\right]\;.
\end{equation}
Therefore we focus on the components of the boson mass matrix $m_B^2$ that are connected to $W_{\rho\tilde \rho \Phi} W^{\dagger \Phi}$.
It is useful to divide the index $a$ into $a_1$ and $a_2$, where $a_1=1,\cdots, n$ and $a_2=n+1,\cdots, N_c$.
Similarly, divide $I$ so that $I=(I_1,I_2)$ such that $I_1=1,\cdots, n$ and $I_2=n+1,\cdots, N$.
Then we can check that those that are coupled to the off-diagonal element $W_{\rho\tilde \rho \Phi} W^{\dagger \Phi}$ can be written as
\begin{equation}\label{E:mB2exp}
{m_B^2}_{\text{effective}} = \left[\begin{pmatrix} A_1 & B_1 \\ B_1 & A_1 \end{pmatrix}\otimes \unit_{n(N_c-n)}\right]\oplus \left[\begin{pmatrix} A_2 & B_2 \\ B_2 & A_2 \end{pmatrix}\otimes \unit_{(N-n)(N_c-n)}\right]\ ,
\end{equation}
where
\begin{equation}
A_1=\bordermatrix{
	&	\Phi^{e_1}_{f_2}		&	\Phi^{e_2}_{f_1}	&	\rho^{M_1}_{e_2}	&	\tilde\rho^{e_2}_{M_1}	&	Z^{M_1}_{e_2}	&\tilde Z^{e_2}_{M_1}\cr
\Phi^{\dagger b_2}_{a_1}	&	\alpha^2\beta^2\mu^2	&	0	&	\mu\alpha\beta\gamma	&	0	&	\alpha^2\beta m\mu	&	0\cr
\Phi^{\dagger b_1}_{a_2}	&	0	&	\frac{\mu^2}{\alpha^2\beta^2}	&	0	&	\frac{\mu\gamma}{\alpha\beta}	&	0	&	\frac{m\mu}{\alpha^2\beta}\cr
\rho^{\dagger a_2}_{I_1}	&	\mu\alpha\beta\gamma	&	0	&	\frac{\mu^2}{\alpha^2\beta^2}+\frac{m^2}{\alpha^2}+\gamma^2	&	0	&	\frac{mm_z}{\alpha}+\alpha\gamma m	&	0\cr
\tilde\rho^{\dagger I_1}_{a_2}	&	0	&	\frac{\mu\gamma}{\alpha\beta}	&	0	&	\alpha^2\beta^2\mu^2+\alpha^2 m^2+\gamma^2	&	0	&	\alpha m m_z+\frac{m\gamma}{\alpha}\cr
Z^{\dagger a_2}_{I_1}	&	\alpha^2\beta\mu m &	0	&	\frac{m m_z}{\alpha}+\alpha \gamma m		&	0	&	\alpha^2 m^2 + m_z^2	&	0\cr
\tilde Z^{\dagger I_1}_{a_2}	&	0	&	\frac{m\mu}{\alpha^2\beta}	&	0	&	\alpha m m_z+\frac{m \gamma}{\alpha}	&	0	&	\frac{m^2}{\alpha^2}+m_z^2
} \ ,
\end{equation}
\begin{equation}\label{E:A2}
A_2=\bordermatrix{
	&	\rho^{M_2}_{e_2}	&	\tilde\rho^{e_2}_{M_2}	&	Z^{M_2}_{e_2}	&\tilde Z^{e_2}_{M_2}\cr
\rho^{\dagger a_2}_{I_2}		&	\frac{m^2}{\alpha^2}+\gamma^2	&	0	&	\frac{m m_z}{\alpha}+\alpha \gamma m	&	0	\cr
\tilde\rho^{\dagger I_2}_{a_2}	&	0				&	\alpha^2 m^2+\gamma^2	&	0	&	\alpha m m_z+\frac{m\gamma}{\alpha}\cr
Z^{\dagger a_2}_{I_2}		&	\frac{m m_z}{\alpha}+\alpha \gamma m	&	0	&	\alpha^2 m^2 + m_z^2	&	0\cr
\tilde Z^{\dagger I_2}_{a_2}	&	0	&	\alpha m m_z+\frac{m\gamma}{\alpha}	&	0	&	\frac{m^2}{\alpha^2} + m_z^2
}\ ,
\end{equation}
\begin{equation}
B_1=\bordermatrix{
	&\Phi^{\dagger f_2}_{e_1}		&	\Phi^{\dagger f_1}_{e_2}	&	\rho^{\dagger e_2}_{M_1}	&	\tilde\rho^{\dagger M_1}_{e_2}	&	Z^{\dagger e_2}_{M_1}	&\tilde Z^{\dagger M_1}_{e_2}\cr
\Phi^{\dagger b_2}_{a_1}	&0	&	0	&	0	&	0	&	0	&	0\cr
\Phi^{\dagger b_1}_{a_2}	&0	&	0	&	0	&	0	&	0	&	0\cr
\rho^{\dagger a_2}_{I_1}	&0	&	0	&	0	&	-\mu^2	&	0	&	0\cr
\tilde\rho^{\dagger I_1}_{a_2}	&0	&	0	&	-\mu^2	&	0	&	0	&	0\cr
Z^{\dagger a_2}_{I_1}&0	&	0	&	0	&	0	&	0	&	0\cr
\tilde Z^{\dagger I_1}_{a_2}	&0	&	0	&	0	&	0	&	0	&	0	
}\ ,
\end{equation}
\begin{equation}
B_2=\bordermatrix{
	&	\rho^{\dagger e_2}_{M_2}	&	\tilde\rho^{\dagger M_2}_{e_2}	&	Z^{\dagger e_2}_{M_2}	&\tilde Z^{\dagger M_2}_{e_2}\cr
\rho^{\dagger a_2}_{I_2}		&0	&	-\mu^2	&	0	&	0\cr
\tilde\rho^{\dagger I_2}_{a_2}	&-\mu^2	&	0	&	0	&	0\cr
Z^{\dagger a_2}_{I_2}		&0	&	0	&	0	&	0\cr
\tilde Z^{\dagger I_2}_{a_2}	&0	&	0	&	0	&	0\cr
}\ ,
\end{equation}
Note that the first summand of the boson matrix matrix \eqref{E:mB2exp} always has a zero eigenvalue whose two eigenvectors are
\begin{equation}
\begin{split}
&\left( \frac{m^2-\gamma \lambda}{m \mu}, 0, \frac{\alpha\beta\lambda} m,0,-\beta ,0 ,0, \frac{m^2-\gamma\lambda}{m\mu},0,\frac{\lambda}{\alpha\beta m},0, -\frac 1 {\beta}\right)\;,\\
&\left( 0,\frac{m^2-\gamma \lambda}{m \mu}, 0, \frac{\lambda}{\alpha\beta m },0,-\frac 1 {\beta} , \frac{m^2-\gamma\lambda}{m\mu},0,\frac{\alpha\beta\lambda}{m},0, -\beta, 0 \right)\;,
\end{split}
\end{equation}
where one is a Goldstone mode and the other a pseudo-modulus.

Using the mass matrices and the formula \eqref{E:CWpotential}, we can compute the effective potential.
Note that the mass matrix expressions $A_1$, $A_2$, $B_1$ and $B_2$ are invariant when $\alpha$ is replaced by $1/\alpha$, or when $\beta$ is replaced by $1/\beta$, up to permutations of rows and columns.
This means that the effective potential \eqref{E:CWpotential} is extremal at $\alpha=\beta=1$.
We only need to check whether it is stable there.
For $\gamma$, there is no such symmetry and the stabilization may occur at a nonzero value of $\gamma$.
A numerical computation has shown that indeed the effective potential is stabilized at $\alpha=\beta=1$ and at some nonzero $\gamma$ as long as $m_z< m$.

\section*{Appendix B\, Semi-local vortex }

In this appendix, we will review basics aspects of a semi-local vortex (see \cite{VAreview} for a review). Usually, a vortex solution is studied by the manifold of minima of a potential $V$. We call the manifold ${\cal M}_{\rm min}$\footnote{One naturally expects that quantum corrections lift the minimum of the potential unless it is not protected by symmetry, so this minimum of the potential generally has the structure $G/H$, where $G$ and $H$ are the symmetry of the theory and that preserved at a vacuum, respectively.}, which can be written as
$${
{\cal M}_{\rm min}={G_{\rm local}\times G_{\rm global}\over H_{\rm local}\times H_{\rm global}}\ .
}$$
For example, in the original Abrikosov-Nielsen-Olesen (ANO) vortex, ${\cal M}_{\rm min}=U(1)$. The theory includes only a $U(1)$ gauge group, which parametrizes the minima of the potential. When $\pi_1({\cal M}_{\rm min})$ is nonzero there should be either a finite energy vortex or an infinite energy one (global vortex). So we have to check whether or not the kinetic term gives a finite contribution to the energy. However, nonzero $\pi_1({\cal M}_{\rm min})$ does not necessarily imply the presence of a finite energy vortex solution when the theory has some global symmetry and vice versa. A vortex generated by symmetry breaking including both global and local symmetry is called a semi-local vortex in a {\it broad sense}. It includes two types of finite energy vortices:
\begin{itemize}
\item The first type is the same as above: $\pi_1({\cal M}_{\rm min})$ is nonzero and the symmetry which supports the vortex is a gauge group. Therefore the kinetic term does not diverge. Conventionally, this type of vortex is called the Abrikosov-Nielsen-Olesen vortex although this is a semi-local vortex in a broad sense. 
\item According to Achucarro, Vachaspati \cite{VA}, even when $\pi_1({\cal M}_{\rm min})=0$, it is possible to create a vortex with finite energy. This type can happen only when both local and global symmetries exist in a theory. This vortex is called a semi-local vortex. Following the convention, we call this type of vortex semi-local.
\end{itemize}
In any case, to have a finite energy solution, the kinetic term of a field should approach zero faster than $1/r$. This condition implies the scalar fields must take values on the same gauge orbit at infinity, and the map from infinity to this gauge orbit defines the winding number $\pi_1(G_{\rm local}/H_{\rm local})$. The finite energy solutions including both ANO and semi-local vortices are classified by this $\pi_1(G_{\rm local}/H_{\rm local})$.

Here we will review the explicit construction of the second type vortex by following the paper \cite{Gibbons} and show a connection to $\pi_2({\cal M}_{\rm moduli})$ in the sigma model limit \cite{Hind}. ${\cal M}_{\rm moduli}$ is called the vacuum moduli and is defined by 
$${
{\cal M}_{\rm moduli}\equiv {{\cal M}_{\rm min}\over L_{\rm orbit}}\ ,\quad {\rm where}\quad L_{\rm orbit}={G_{\rm local}\over H_{\rm local}}\ .
}$$
where $L_{\rm orbit}$ is the gauge orbit in ${\cal M}_{\rm min}$. The model is a $U(1)$ gauge theory with a Lagrangian
\begin{equation}
{\cal L} = - \frac{1}{2} (D_{\mu} \Phi)^\dagger D^\mu \Phi 
-  \frac{\lambda}{8}(\Phi^\dagger \Phi - \eta^2)^2
-\frac{1}{4} F_{\mu \nu} F^{\mu \nu} \ ,
\end{equation}
where $\Phi$ is an $SU(2)$ doublet and the Lagrangian is invariant under the $SU(2)$ global symmetry. The topology of the vacuum manifold is ${\bf S}^3$. Since $\pi_1({\bf S}^3)=0$, one naively expects that there is no stable vortex solution. However, one can construct a stable vortex. Since we are interested in a static solution, the energy density is given by
\begin{equation}
{\cal E} = \frac{1}{2} |D_i \Phi|^2 + \frac{\lambda}{8} (\Phi^\dagger \Phi - \eta^2)^2 + B_i^2\ . \label{Hamiltonian}
\end{equation}
To keep the energy finite, the energy density at infinity must be equal to zero. This implies
\begin{itemize}
\item The scalar takes values in the minimum of the potential, ${\bf S}^3$.
\item The first term in the energy density also vanishes. This confines the vev of the scalar fields in the gauge orbit ${\bf S}^1$ of the local $U(1)$ symmetry in ${\bf S}^3$.
\item The gauge field must be a pure gauge and should be taken so that $D_i \Phi$ vanishes.
\end{itemize}
Therefore, if one identify all the points in the same gauge orbit, the scalar fields take the same value in any direction at spatial infinity. Namely we regard ${\bf S}^1_{\infty}$ as a point. Roughly, this means the two-dimensional space transverse to the vortex is identified as ${\bf S}^2$. Now we can consider a map from this spatial ${\bf S}^2$ to space ${\bf S}^3/U(1) = {\bf S}^2$, or more generally, ${G_{\rm global}/H_{\rm global}}=SU(2)_{\rm global}/U(1)_{\rm diag} = {\bf S}^2$. This is classified by $\pi_2(SU(2)_{\rm global}/U(1)_{\rm diag})$. If a configuration is trivial under this $\pi_2$, then we can take a scalar value in spatial ${\bf CP}^1$. Thus, without causing a divergence of the kinetic term one can change the configuration to a trivial one\footnote{Note that if we weakly gauged $SU(2)_{\rm global}$, then the configuration is not stable anymore. }. It is worth noting the correspondence between $\pi_1(G_{\rm local}/H_{\rm local})$ and  $\pi_2({\cal M}_{\rm mod!
 uli})$. The semi-local vortex considered in Gibbons et al. is a one-parameter generalization of the ``usual" vortex solution. By ``usual", we mean that the scalar fields take values in the same gauge orbit everywhere on $R^2$ transverse to the vortex. In this sense, the usual vortex is some limit of a semi-local vortex if there exists a one-parameter generalization. On the other hand, semilocal vortices can be understood as lump solutions in non-linear sigma models obtained in the low-energy limit and whose target space is the vacuum moduli. These lump solutions can be classified by $\pi_2 (G_{\rm global}/H_{\rm global})$, so it should agree with $\pi_1(G_{\rm local}/H_{\rm local})$---they both classify semi-local vortices
in different limits. Abrikosov-Nielsen-Olesen vortex is an exception, because it does not have any one-parameter generalization.

\section*{Appendix C\, D-brane and cosmic strings}

\begin{figure}[htbp]
\begin{center}
  \epsfxsize=10cm \epsfbox{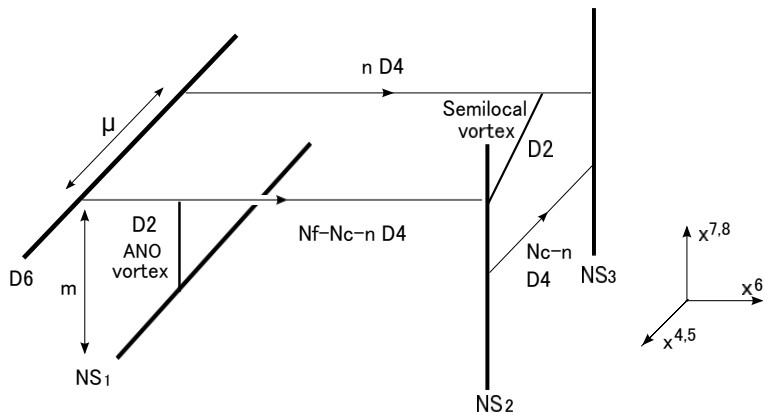}
\end{center}
 \vspace{-.5cm}
\caption{\sl Two kinds of vortices. One is Abrikosov-Nielsen-Olesen vortex which has tension $\mu_T \sim m^2$. It is D2 brane extending vertical direction. On the other hand, horizontal D2 brane is semi-local vortex. The tension is $\mu_T\sim \mu^2$.  For precise argument one of the NS branes in the vertical direction should be replaced D6 for semi-local vortex.  }\label{Brane}
\end{figure}

In \cite{KOO}, D-brane configurations are presented for all supersymmetry breaking vacua in the model studied in section 2 in the main text. In this appendix we would like to identify brane configurations for ANO and semi-local vortices. We start with a review of a magnetic brane configuration shown in \cite{KOO} and its slight modification for our purpose.  Consider Type IIA superstring theory in 
the flat 10-dimensional Minkowski spacetime with coordinates
$x^{0,\cdots,9}$.  Introduce three NS5 branes. The first one, which we call NS$_1$, is located at $x^{7,8,9}=0$ and extended in the $x^{0,\cdots,3}$ and $x^{4,5}$ directions ($z=x^4+ix^5$ in complex coordinates.). The other two are going in directions $w=x^7+ix^8$ and $x^{4,5,9}=0$ but placed at different points in $x^6$. We call them NS$_2$, NS$_3$. Also we introduce a D6 brane extending in $x^{4,5,9}$ directions and placed at $x^{7,8}=0$. For lower energy configurations, we then suspend $(N_f-N_c-n)$ D4 branes between
D6 and NS5$_2$, $N_C-n$ D4 branes between NS$_2$ and NS$_3$, and
$n$ D4 branes between D6 and NS$_3$. The total intersecting brane configuration is shown in Figure \ref{Brane}.

Before identifying a vortex configuration in this D-brane system, let us comment on the gaugino mass and tachyonic direction from the point of view of D-branes. Suppose we embed the SM group into $SU(N_f-N_c-n)$. This open string mode connecting between $N_f-N_c-n$ and $N_c-n$ corresponds to a messenger. When the distance between these D4 brane becomes very close, a tachyonic mode is developed and it causes a reconnection of D4 branes. As was shown in the section 2, in this case, the leading order of gaugino masses are non-zero. 

On the other hand, if the SM group is embedded into $SU(n)$, the messenger, which is open string connecting two $n$ D4 and $N_c-n$ D4 branes, can never be tachyonic. In this case, the leading order gaugino masses are vanishing. This fact is consistent with argument shown in \cite{KS}.

In a supersymmetric configuration, ANO and semi-local vortices were already studied by Hanany and Tong \cite{Tong} (see \cite{TongII} for review). They claimed that a soliton with co-dimension two is $k$ D2 branes which are suspended between a NS five brane and D4 branes. The world-volume of the D2 branes is extending in $x^{0,1,2}$. We can identify two kinds of such D2-branes. One is suspended vertically connected to NS$_1$ brane. Since extra D4 branes are not connected to the NS$_1$ brane, there is no extra degree of freedom. On the other hand, D2 branes going in the horizontal direction can have an extra-degree of freedom since $N_f-N_c-n$ D4 branes are connected to the NS$_2$ brane. We identify this vertex as a semi-local vortex and the other D2 branes are ANO vortex which we found in the main text. The tension which is proportional to the length of the D2 branes is consistent with the symmetry breaking scale of the group under which the vortex is generated. 

Finally, we discuss the decay of semi-local vortices. As reviewed in Appendix B, a semi-local vortex becomes unstable when one gauges global symmetries. In the brane configuration, gauging of non-abelian groups in metastable vacua corresponds to the replacement of the D6 brane with a NS brane. Once we replace it, a gauge field on D4 brane is introduced, so one may say that the end point of D2 can be moved freely and can shrink to zero size. On the other hand, if there is no gauge field on D4, moving the end of D2 branes costs infinite energy. So it cannot move and the vortex cannot shrink. It would be interesting to explore further on this decay process and study the divergence of the kinetic term from the D-brane point of view.

%
%
%

\end{document}